\documentclass[aps,prx,twocolumn,superscriptaddress,10pt]{revtex4-2}
\usepackage{graphicx} 
\usepackage{amsmath,amssymb}
\usepackage{physics}
\usepackage{booktabs}
\usepackage{graphicx}
\usepackage[table]{xcolor} 
\newcommand{\lightrule}{%
  \arrayrulecolor{black!40}\specialrule{0.3pt}{0pt}{0pt}\arrayrulecolor{black}%
}
\usepackage{comment}
\usepackage{hyperref}
\hypersetup{
    colorlinks=true,
    linkcolor=blue,  
    citecolor=blue,
    urlcolor=magenta,
    }

\def \z2sf {$\mathbb{Z}_2^{\rm sf}$ }

\begin{document}

\title{Icosahedral discretizations of the state-operator correspondence of \texorpdfstring{$d=3$}{d3} conformal field theories}
\author{Hansen S. Wu}
\affiliation{Department of Physics, The Pennsylvania State University, University Park, PA -16802.}

\author{Ribhu K. Kaul}
\affiliation{Department of Physics, The Pennsylvania State University, University Park, PA -16802.}
\affiliation{Center for Theory of Emergent Quantum Matter, The Pennsylvania State University, University Park, PA -16802.}

\begin{abstract}
We propose that scaling dimensions of $d=3$ conformal field theories can be studied on a system of qubits with near term quantum simulation platforms. Our proposal chooses couplings of quantum many-body problems on a polyhedral lattice at which the conformal state-operator correspondence can be observed most accurately in the spectrum. We validate our protocol on the Ising model where we extract the scaling dimensions of a number of scaling operators with a few percent accuracy from the spectrum of a system of just 20 qubits. The procedure makes only minor assumptions beyond general conformal invariance -- it may hence be applied widely.  Requirements and challenges to realize this proposal on quantum computers are discussed. Our results demonstrate that for current or near term qubit platforms, three dimensional conformal field theories present a unique opportunity -- a forefront problem that is difficult on classical computers but may be solved through quantum simulation. 
\end{abstract}
\date{\today}
\maketitle

\section{ Introduction} 
The study of critical phenomena has been both a challenging and enduring pursuit in theoretical physics \cite{Cardy_1996}. It is well known that the nature of critical phenomena depends drastically on the space time dimension of the system. While two dimensional criticality is rich, it is now well understood because of the vast array of specialized methods by which it can be studied \cite{DiFrancesco:1997nk, ginsparg1988appliedconformalfieldtheory}.  The sophisticated methods used in $d=2$ do not generalize easily to higher dimensions.  On the other hand, many of the paradigmatic examples of critical phenomena are known to simplify in four and higher dimensions \cite{RevModPhys.55.583}. All this makes three dimensions the enigmatic case which is both intriguing and relatively poorly understood -- it is on this case of $d=3$ that we will focus here. 

One central point of interest in critical phenomena is the emergence of a scale invariant continuum description, which is characterized by various universal pure numbers that transcend any microscopic realization of the critical point~\cite{Cardy_1996}. The most fundamental of these pure numbers are the dimensions of scaling operators. 
Historically the scaling dimensions of three dimensional critical field theories have been studied by approximate (semi)-analytic methods on continuum theories and Monte Carlo simulations of lattice models making use of scale invariance (see e.g.~\cite{RevModPhys.55.583, PELISSETTO2002549, PhysRevB.82.174433}).  It has been conjectured and is now widely accepted that critical points in $d=3$ also exhibit conformal invariance \cite{Polyakov:1970xd, Rychkov_2017}. Taking advantage of this higher symmetry, in the last decade major progress in the $d=3$ conformal bootstrap method has made accurate estimates of operator scaling dimensions \cite{PhysRevD.86.025022, Poland_2019, Rychkov_2025}. In another recent development, the fuzzy sphere regularization \cite{Zhu_2023} has demonstrated remarkable success in taking advantage of the state-operator correspondence to study scaling dimensions. While both are landmark developments, they each have to make their own assumptions and have specific limitations. It is hence of great interest to develop alternate methods to study scaling dimensions of $d=3$ conformal field theories. 

\begin{figure}
    \centering
    \includegraphics[width=1.0\linewidth]{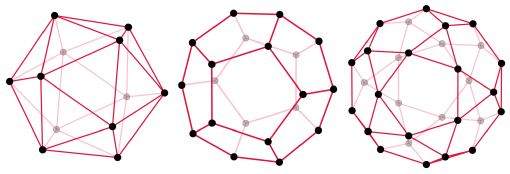}
    \caption{In the manuscript we validate our proposal on two platonic polyhedral solids: the icosahedron and the dodecahedron. A third solid the icosidodecahedron (which is not a platonic solid but has site and edge transitivity) with 30 sites has also been studied here and its spectrum is discussed in App.~\ref{app:idodec}. } The dodecahedron is the dual of the icosahedron with faces and vertices interchanged. We place qubits on the vertices (resulting in a Hilbert space of $2^{12}$ and $2^{20}$) and choose interactions to realize quantum models with an Ising \z2sf symmetry with a conformal spectrum. Even though both structures share the same icosahedral $I_h$ symmetry (the largest discrete subgroup of $O(3)$), we present evidence that the dodecahedron has a  more accurate conformal spectrum than the icosahedron, which we attribute to the increase of Hilbert space dimension. Increasing the Hilbert space dimension further preserving $I_h$ should yield even more accuracy -- although out of reach of exact diagonalization on a classical computer, it is proposed as an ideal project for near term quantum simulators.
    \label{fig:id}
\end{figure}

In a completely different development, in the last decade quantum simulators have appeared as an exciting new technological paradigm with various complementary platforms under intense study \cite{feynman2018simulating, Georgescu_2014, PRXQuantum.2.017003}. For the purposes of this work, we will view the simulators as a system of qubits in which Hamiltonian time evolution can be programmed in a controlled way and then qubits measured, allowing access to features of quantum many-body physics.  While the current systems typically have less than about 100 qubits, the physical processes of these qubits are already approaching a complexity which is beyond the reach of classical simulation \cite{arute2019quantum, kim2023evidence}, enabling a new window to basic models of quantum many-body physics, such as the transverse field Ising Hamiltonian \cite{Bernien_2017, Ebadi_2021, RevModPhys.93.025001, PhysRevX.5.021027}. An exciting contemporary challenge then is to identify problems of physical interest for which quantum simulators can provide us solutions that are out of reach for classical computers. In this manuscript we discuss how quantum computers with only 100 or so qubits typical to current or near term quantum simulation with sufficient control can obtain universal numbers of conformal field theories such as the dimensions of scaling operators, with an accuracy out of reach of classical computers. We validate our proposal for the paradigmatic $d=3$ Ising conformal field theory, for which we find that with a system of just 20 qubits (which we can study with classical simulations), we get results accurate to a few percent. 

\begin{figure}[htbp]	
\centering
\setlength{\abovecaptionskip}{20pt}
\includegraphics[angle=0,width=0.45\textwidth]{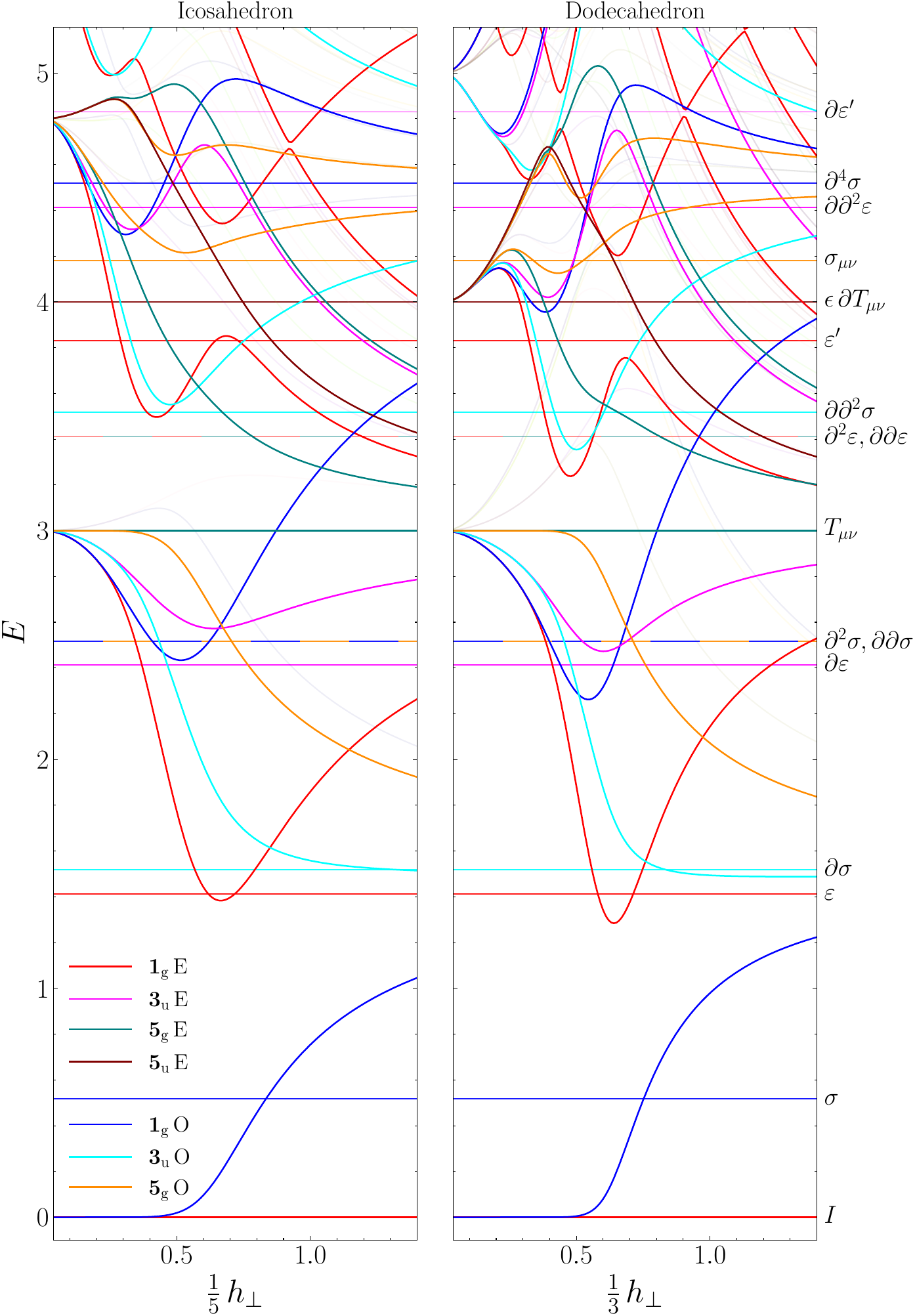}
\caption{The spectrum of the nearest neighbor transverse field Ising model Eq.~\ref{eq:tfim} on the icosahedron (12 qubits) and dodecahedron (20 qubits) as a function of $h_\perp$. The spectrum of the TFIM on the icosahedron (left) and on the dodecahedron (right) has appeared previously~\cite{Lao_2023,cruz2026:yl}. We  divide $h_\perp$ by the coordination number (3 and 5), so that we may view both spectra on the same scale. We have subtracted the ground state energy from each level at each coupling. The ground state is hence by definition at exactly zero energy. It is in the $\mathbf{1}_g$E irrep consistent with the identity operator, $I$.  We have restricted ourselves to the 7 irreps of the TFIM on these $I_h$ symmetric graphs shown in the legend, because these sectors contain all the low lying energy levels of interest here. In light gray we show for reference energy levels from other sectors. At each $h_\perp$ we have rescaled the energy so the lowest level in the $\mathbf{5}_g$E sector is 3, since we identify this level with the stress-tensor $T_{\mu\nu}$. Also shown as straight lines and labeled on the right edge of the right panel  are the bootstrap values for the $\Delta$ of each of the low-lying scaling fields. The colors of the bootstrap lines are chosen to match the expected representations of the $I_h\times$\z2sf for these levels. A line with alternating colors indicates that CFT predicts two operators with different quantum numbers that have the same $\Delta$. Clearly there is a range of $h_\perp$ at which the bootstrap and icosahedral spectrum results seem close for both solids. A careful look at this picture already seems to show that this matching is better on the dodecahedron than the icosahedron (even though they share the same symmetry) apparently because the dodecahedron has a larger Hilbert space. We explore this question quantitatively further in this section.  }
\label{fig:id_spec}
\end{figure}

\section{ Icosahedral Spectroscopy} 
\label{sec:is}
The state operator correspondence implies that the scaling dimensions ($\Delta$) of a CFT are equal to gaps in the spectrum of certain quantum problems defined on $S^{d-1}$ \cite{Cardy:1984rp, Cardy:1985lth}. This has been used extensively for (1+1)-$d$ dimensional CFT, where the $S^1$ geometry can be conveniently discretized using the quantum  systems on a chain lattice with periodic boundary conditions \cite{CARDY1986186}. For (2+1)-$d$ CFTs, the quantum many-body physics problem needs to be put on a sphere, but there is no sequence of discrete subgroups that approach $O(3)$. Indeed the largest discrete subgroup of $O(3)$ is $I_h$ (full icosahedral group) \cite{cohan1958spherical, Brower_2013}. In (2+1)-$d$, it has been argued that the finite size spectrum in the torus geometry is universal, but the scaling dimensions are not known to be related to the gaps as they are for $S^2$~\cite{PhysRevLett.117.210401}. A novel idea to get around this difficulty is the fuzzy sphere which discretizes the quantum many-body problem while preserving $O(3)$ symmetry~\cite{Zhu_2023}. However, there is no known way to build the fuzzy sphere problem from qubits and as such it is not amenable to quantum simulation. Here we show that ``icosahedral spectroscopy"  (spectrum of $I_h$ symmetric structures) is an alternate route to produce accurate results. While it has the drawback of breaking $O(3)$, it has the advantage of being naturally formulated with qubits.

 The most popular way to regularize the $d=3$ Ising conformal field theory as a quantum many-body problem is using the two-dimensional transverse field Ising model (TFIM)
\begin{equation}
\label{eq:tfim}
    H_{\rm TFIM} = -\sum_{\langle ij \rangle} Z_i Z_j - h_\perp \sum_i X_i
\end{equation}
and then tuning the ground state so that $h_\perp$ is at the critical point \cite{Suzuki:1976ldt,sachdev2011}. For the purpose of quantum simulation we may view this system as being built from a set of qubits. The model realizes the all important spin flip symmetry \z2sf microscopically by the operator $U_{\rm sf}= \Pi_i X_i $. In the RG picture there is only one relevant direction that is \z2sf  even, which is tuned to zero by $h_\perp$, all other directions are irrelevant and their couplings go to zero, resulting in the universal conformal field theory \cite{Cardy_1996}.
To use the state operator correspondence for a $d=3$ CFT in a lattice formulation, we will approximate $S^2$ by a polyhedral solid \cite{Cardy:1985lth, alcaraz1987numerical, Lao_2023}. The idea of using a polyhedral TFIM quantum Hamiltonian for the state-operator correspondence was proposed in the original work of Cardy~\cite{Cardy:1985lth}, but appears to have received only limited attention~\cite{alcaraz1987numerical,Lao_2023,cruz2026:yl}. We note that a related sequence of works has used Cardy's idea to study the $d=3$ Ising CFT from classical statistical mechanics of Platonic polyhedral structures \cite{Weigel_2000, Brower_2013}. Because of its natural realization on a quantum computer, we shall focus here on the quantum Hamiltonian perspective, which was most thoroughly considered in Ref.~\cite{Lao_2023} who studied the spectrum of the TFIM on the icosahedron (12 sites) and found that using conformal perturbation theory (which makes use of some known conformal data of the Ising CFT) the spectrum is consistent with the Ising CFT perturbed with a small number of operators~\cite{Lao_2023} (the 12 qubit icosahedron was also used recently for quantities beyond scaling dimensions~\cite{qcp6-v1dr}). In this section we show that by tuning two couplings in an unbiased way and studying a larger structure (dodecahedron) we can determine the primary $\Delta$ accurately to a few percent from 20 qubits without making use of any Ising specfic conformal data.

The sphere has $O(3)$ symmetry (including parity). The largest discrete subgroup of $O(3)$ is the icosahedral group $I_h$ --  we only consider structures with this symmetry. The simplest solids with this symmetry is the icosahedron (12 vertices), and the dodecahedron (20 vertices), both of which are edge, face and vertex transitive, Fig.~\ref{fig:id}.   We shall focus on these two solids in our study here and show that we can already get reasonably accurate numbers. While $I_h$ symmetric structures with more vertices exist, they are out of reach of our ED on classical computers, but we propose they are amenable to near term quantum simulation - we discuss them in Sec.~\ref{sec:qs}. 

For any icosahedrally symmetric structure we may classify the quantum many-body energy levels according to which irreducible representation of the proper icosahedral rotations they transform under $(\mathbf{1},\mathbf{3},\mathbf{3}^\prime,\mathbf{4},\mathbf{5})$.  In addition, since both the icosahedron and dodecahedron structures have inversion symmetry  we label the energy labels by ``g” or ``u”(gerade/ungerade). Table~\ref{tab:o3_Ih} shows the $O(3)$ contributions to each icosahedral sector.  At low energies, we can unambiguously distinguish states up to $\ell = 2$ by identifying $\ell = 0 \rightarrow \mathbf{1}_{\mathrm{g}}$, $\ell = 1 \rightarrow \mathbf{3}_{\mathrm{u}}$, and $\ell = 2 \rightarrow \mathbf{5}_{\mathrm{g}}$. The first higher $\ell$ contamination occurs in sector $\mathbf{5}_{\mathrm{g}}$ from $\ell = 4$, blurring the distinction between $\ell = 2$ and $\ell = 4$ states in the $\mathbf{5}_{\mathrm{g}}$ sector at energies around $\sim 4$. Moreover, the $\ell = 3$ sector splits into $\mathbf{3}'_{\mathrm{u}}\oplus\mathbf{4}_{\mathrm{u}}$ preventing identification of states of $\ell \geq 3$. Despite these challenges, access to $\ell \in \{ 0,1,2 \}$ for low energies allow us to label the most interesting portion of the conformal spectrum and to observe the conformal multiplet structure. Finally since we are studying the TFIM with \z2sf symmetry we can add another quantum number for whether the levels are spin flip ``E" or ``O" (even/odd). Since the proper rotations, inversion and spin flip all commute with each other, taken together the Hamiltonian and hence its eignspectrum breaks up into $5\times 2 \times 2 =20$ symmetry sectors.

\begin{table}[!t]
\centering
\resizebox{0.7\linewidth}{!}{%
\begingroup
\setlength{\tabcolsep}{7pt}
\renewcommand{\arraystretch}{1.15}
\begin{tabular}{l|ccccccc}
\toprule
\multicolumn{1}{c}{$\ell$} & 0 & 1 & 2 & 3 & 4 & 5 & 6 \\
\midrule
$\mathbf{1}$   & 1 &   &   &   &   &   & 1 \\
$\mathbf{3}$   &   & 1 &   &   &   & 1 & 1 \\
$\mathbf{3}'$  &   &   &   & 1 &   & 1 &   \\
$\mathbf{4}$   &   &   &   & 1 & 1 &   & 1 \\
$\mathbf{5}$   &   &   & 1 &   & 1 & 1 & 1 \\
\bottomrule
\end{tabular}
\endgroup
}
\caption{Embedding of $O(3)$ inside $I_h$, adapted from \cite{cohan1958spherical}}
\label{tab:o3_Ih}
\end{table}

In Fig.~\ref{fig:id_spec} we show the low lying energy levels of the TFIM on the icosahedron (left) and the dodecahedron (right) as a function of $h_\perp$.  To prevent the plot from getting too busy, we have chosen 8 of the 20 symmetry sectors (see legend) that are most significant for our purposes (others are shown in gray). The energy has been rescaled at each coupling so that the lowest ${\bf 5}_{\mathrm g}$E, which is expected to correspond to the stress tensor is fixed to 3~\cite{Zhu_2023}. Also shown are the conformal bootstrap scaling dimensions colored with the expected $I_h$ irrep. As discussed in the caption at a coupling close to $\sim$0.8, we observe an approximate matching of the conformal bootstrap with the icosahedral spectroscopy, with qualitative evidence that the dodecahedron is closer to the ideal conformal spectrum than the icosahedron.  In the remainder of this section our goal is to make this observation quantitative: First, we study a procedure using conformal spectral constraints to tune microscopic couplings on a finite system to a special point where the conformal spectra can be observed. As we shall explain below this method only assumes conformal invariance and a very plausible identification of energy states with operators  based on their icosahedral quantum numbers. Importantly, it requires no prior knowledge of the primary scaling dimensions. Secondly, we present compelling evidence that the quality of the conformal spectra improves when going from the icosahedron to the dodecahedron, which share the same $I_h$ symmetry but differ in the number of qubits from 12 to 20 (the Hilbert space increases from $2^{12}=4028$ to $2^{20}=1048576$). This leads us to the natural conjecture that increasing the number of qubits preserving $I_h$ can lead to better conformal spectra, despite the spatial symmetry remaining the same.

\begin{figure}[!t]
    \centering
    \includegraphics[width=1.0\linewidth]{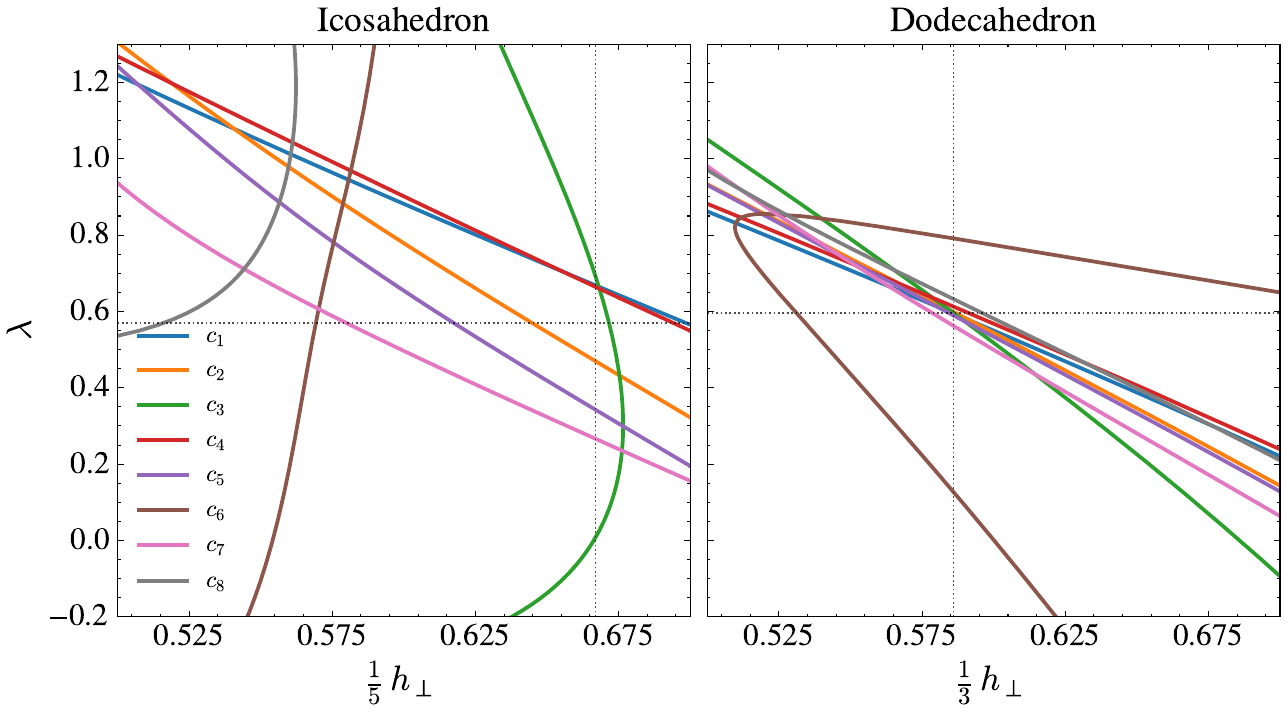}
    \caption{Constraint curves satisfying $c_{i}(h_{\perp},\lambda)=0$ with each $c_i$ defined in Eqn.~\ref{eq:constrain_eqns}. As the parameters have been rescaled to be comparable across graphs, the observation that the constraint curves have less spread for the dodecahedron is evidence that the conformal point is better defined for a system with more qubits. Crosshairs mark the minimization of $|c|$ for the $\sigma$-tower of constraints.}
    \label{fig:constraint_curves}
\end{figure}

The structure of conformal invariance implies that the eigenvalues on the sphere are organized into towers consisting of primary operators and their descendants. It is well known that if the scaling dimension of the primary field is $\Delta$, the  descendants at level $k$ have scaling dimension $\Delta+k$~\cite{Rychkov_2017}. We may use this structure to identify the microscopic couplings which best describe the conformal field theory (similar ideas were used in the context of the fuzzy sphere~\cite{xstj-xvcy,zhou2025fuzzifiedjuliapackage}). Thus without making any assumptions on the scaling dimensions of the primary operators, we know that descendant operators must be spaced by integer energy gaps (once the speed of light is fixed, as we have by setting $E_{T_{\mu\nu}}=3$). There are many such constraints, with the eight lowest in energy being,
\begin{equation}
\label{eq:constrain_eqns}
\begin{aligned}
    c_1 &= E_{\partial \sigma} - E_{\sigma}-1\\
    c_2 &= E_{\partial\partial \sigma} - E_{\sigma} - 2\\
    c_3 &= E_{\partial^2 \sigma} - E_{\partial\partial\sigma}\\
    c_4 &= E_{\partial\varepsilon} - E_{\varepsilon}-1\\
    c_5 &= E_{\partial\partial \varepsilon} - E_{\varepsilon} - 2\\
    c_6 &= E_{\partial^2 \varepsilon } - E_{\partial\partial\varepsilon}\\
    c_7 &= E_{\epsilon\partial T} - E_{T} - 1\\
    c_8 &= E_{\partial\varepsilon'} - E_{\varepsilon'} - 1,
\end{aligned}
\end{equation}
which are satisfied when $c_{i}=0$. With one tuning parameter, each constraint can only be satisfied at isolated points (if at all). However, adding a second parameter yields curves in parameter space satisfying $c_{i}=0$ for each constraint. To study this for the TFIM Hamiltonian, we choose to add the coupling $-\frac{\lambda}{2}\sum_{\langle ij\rangle}(X_i X_j + Y_iY_j)$ which maintains the \z2sf symmetry.  For each of the eight constraints, we locate the curves $c_{i}(h_{\perp},\lambda) = 0$, which are shown in Figure~\ref{fig:constraint_curves} for a zoomed in region of the microscopic couplings $(h_\perp,\lambda)$. Two constraint curves can cross at a point. For an ideal conformal spectrum, we would expect all pairs of 8 curves to cross at the same common point. We observe that the crossing points vary only slightly between pairs of constraints, which we attribute to finite size effects. Confirming this hypothesis, we find that while the crossing points of the constraint curves are more broadly spread for the icosahedron, those of the dodecahedron exhibit a smaller region where many curves converge (we study the same window of couplings). This is the first piece of evidence that the dodecahedron is an improvement as the conformal point is better defined.

To turn these constraints into an estimate of the critical point (and whence  the $\Delta$), we minimize $|c|^2=\sum_{i\in S} c_{i}^2 $ over a chosen subset $S$ of the eight constraints using the Gauss-Newton method \cite{nocedal2006numerical}.  The crosshairs in Figure~\ref{fig:constraint_curves} identify the point where $|c|$ is minimized with respect to the $\sigma$ tower of constraints  $S_{\sigma}=\{1,2,3\}$. In Table~\ref{tab:scaling_dimensions}, the most relevant scaling dimensions extracted on the dodecahedron vary at the few percent level on which set of constraints we try to satisfy. The variation is higher for the icosahedron than the dodecahedron, leading us to conclude that these variations are a finite size effect that would get smaller as the size of the Hilbert space is increased. Of the constraint sets listed, $S_{\sigma}$ or $S_{\partial\mathcal{O}}$ on the dodecahedron yields scaling dimensions closest to the bootstrap values. It is clear from either Table~\ref{tab:scaling_dimensions} or Figure~\ref{fig:constraint_curves} that the dodecahedron hosts a better conformal point than the icosahedron. Despite having the same spatial symmetry, it seems that simply increasing the number of sites improves the conformal spectrum. While our model only has $I_h$ symmetry, the CFT spectrum should have full $O(3)$ symmetry. This would require some accidental degeneracies in our spectrum {e.g.} This implies a spectrum where ${\bf 3^\prime}$ and ${\bf 4}$ merge into a single 7 fold degenerate $\ell=3$ level. Such $\ell=3$ levels  would appear as descendants e.g. third descendant of a scalar primary, or as some primary of very high scaling dimension. We note that we do not observe this merging in our spectrum even at the finely tuned point (the ${\bf 3^\prime}$ and ${\bf 4}$ are split). This indicates that we have not achieved full $O$(3) symmetry, even so the numerical values of the scaling dimensions are reasonably accurate.

\begin{figure}[t!]	
\centering
\setlength{\abovecaptionskip}{20pt}
\includegraphics[angle=0,width=0.45\textwidth]{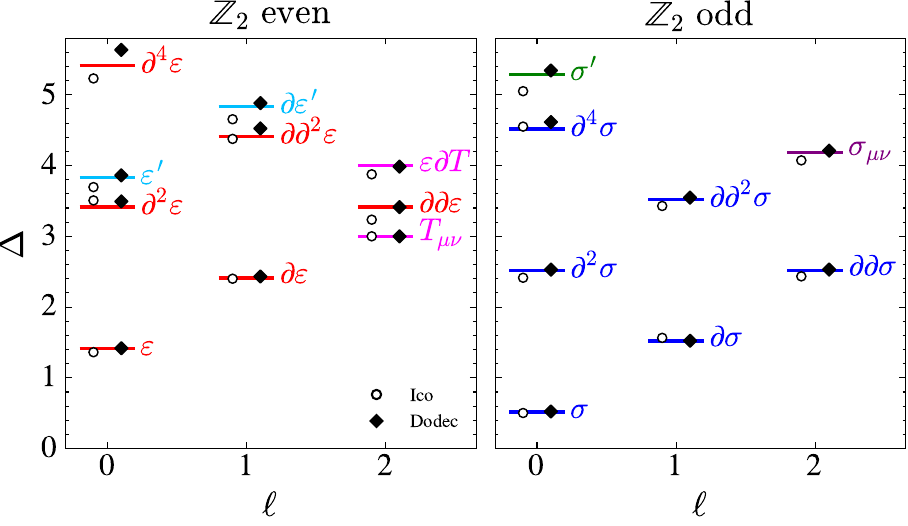}
\label{spectrum}
\caption{Comparison of scaling dimensions and spin inferred from our study on the icosahedron (open circles) and dodecahedron (solid circles) with the conformal bootstrap (solid lines) for the \z2sf even (left) and odd (right) sectors. The spectrum here is shown for lattice couplings $(h_\perp,\lambda)$ at which the $\sigma$-tower constraints ($c_1,c_2,c_3$) are best satisfied according to the Gauss-Newton method discussed in the text.}
\end{figure}

\begin{table}[t]
    \centering
    \begin{tabular}{l|ccccc}
    \toprule
    $\mathcal{O}$ & $\sigma$ & $\varepsilon$ & $\varepsilon'$ & $\sigma_{\mu\nu}$ & $\sigma'$ \\
    \midrule
    Bootstrap            & 0.518 & 1.413 & 3.830 & 4.180 & 5.291 \\
    Fuzzy sphere         & 0.524 & 1.414 & 3.838 & 4.214 & 5.303 \\
    \lightrule
    Dodec, $\sigma$      & 0.526 & 1.420 & 3.860 & 4.210 & 5.342 \\
    Ico, $\sigma$  & 0.503 & 1.362 & 3.695 & 4.073 & 5.051 \\
    \lightrule
    Dodec, $\varepsilon$      & 0.554 & 1.402 & 3.658 & 4.137 & 5.003 \\    
    Ico, $\varepsilon$  & 0.499 & 1.256 & 3.539 & 3.962 & 4.520 \\
    \lightrule
    Dodec, $\partial\mathcal{O}$      & 0.532 & 1.416 & 3.822 & 4.197 & 5.278 \\    
    Ico, $\partial\mathcal{O}$   & 0.543 & 1.257 & 3.506 & 3.886 & 4.370 \\
    \lightrule
    Dodec, all      & 0.543 & 1.424 & 3.804 & 4.189 & 5.242 \\    
    Ico, all  & 0.520 & 1.313 & 3.548 & 4.002 & 4.662 \\

    \bottomrule
    \end{tabular}
    \caption{Conformal dimensions from bootstrap and fuzzy sphere (adapted from \cite{Zhu_2023}) compared to the icosahedron and dodecahderon results from this work. For our results we have shown the dimensions implementing four different sets of constraints the $\sigma$-tower constraints $S_{\sigma}=\{1,2,3\}$, the $\varepsilon$-tower constraints $S_{\varepsilon}=\{4,5,6\}$, the first descendant constraints $S_{\partial\mathcal{O}}=\{1,4,7,8\}$ and finally all eight constraints in Eq.~\ref{eq:constrain_eqns}. The variations in the scaling dimensions across the different sets of constraints is smaller in the dodecahedron than the icosahedron. }
    \label{tab:scaling_dimensions}
\end{table}

\section{ Quantum Simulation} 
\label{sec:qs}

Before discussing the suitability of our setup to quantum simulation we discuss the possibility of using classical computing methods to study this problem. There are three well known unbiased methods to study such systems and each of them will find this problem challenging. Here we employed exact diagonalization which cannot be extended for more than 40 or so qubits because of the exponential growth of the Hilbert space. The density matrix renormalization group is most powerful in one dimensional quantum systems and suffers from convergence and scaling when extended to higher dimensions \cite{Stoudenmire2012}. Finally, quantum Monte Carlo is the most promising since the models we have discussed so far do not suffer from the sign problem. Even so, 
extracting the spectrum of excitation energies from imaginary time data is an extremely challenging problem on which there has been only limited progress (see e.g.~\cite{gtbl-vsdt} for a recent work). 

In the previous section, we have seen how the $\Delta$ of the five lightest (excluding $I$ and $T_{\mu\nu}$ which we set by fiat to 0 and 3, respectively) conformal primaries  of the $d=3$ Ising field theory can be inferred to a few percent accuracy from the spectrum of a quantum many-body problem realized on a system of 20 qubits well within reach of current technologies. We now discuss the general requirements that a qubit platform for quantum simulation would need to satisfy to implement this proposal. Since the technology is rapidly developing, we focus on general requirements rather than specific physical platforms.  However, for each requirement, we note experimental studies on a hardware that have demonstrated required (or closely related) capabilities.  

{\bf Number of Qubits:} The relatively modest number of qubits discussed makes our proposal well within the capabilities of a number of current qubit platforms, including Rydberg atoms, trapped ions and transmons \cite{PRXQuantum.2.017003}.  An important point of discussion is how our proposal would scale up to improve accuracy with more qubits. Clearly we need $I_h$ symmetric polyhedral structures with larger Hilbert spaces than those studied in this paper -- these are mostly out of reach of ED on classical computers but are of interest for future quantum simulation. If we require vertex, edge and face transitivity (``platonic solids"), the icosahedron and the dodecahedron are the only polyhedra \cite{cromwell1997polyhedra}. To further increase the size of the Hilbert space (maintaining the $I_h$ symmetry), there are at least two different approaches. {\bf Symmetric solids:} If we keep vertex transitivity, but relax edge and face transitivity we can study quantum models on Archimedean lattices (up to 120 vertices). Also available are their duals -- Catalan solids (up to 92 vertices), which have face transitivity only. While this approach of highly symmetric structures still has a maximum Hilbert space, given the improvement we have seen from 12 to 20 qubits, it is conceivable that these structures will be large enough for practical purposes. {\bf Icosahedral Refinement:} A second approach to go beyond platonic solids is to triangulate the sphere by dividing the faces of a icosahedron into smaller triangles (at each level of refinement one triangle is replaced by four triangles by inscribing a triangle in the larger triangle, the simplest sequence of such structures would have a number of qubits, $10s+2$ with $s\in \mathbb{Z}^{+}$ see e.g. \cite{brower2024isingmodelmathbbs2}). While this procedure does not preserve vertex, edge or face transitivity, it can be refined with no limit on the number of qubits and allows for an arbitrary increase of the Hilbert space as a dense covering of the sphere is approached. We reiterate that increasing the number of qubits cannot increase the symmetry of the structure closer to $O$(3), but with fine tuning it is hoped that it would be possible to get closer to the Ising fixed point even with this lower symmetry. We do not have a procedure to do this systematically yet -- it is an interesting direction for future research.

{\bf Icosahedral geometry:} We need the qubits to interact on a three dimensional polyhedral geometry.  This may be achieved through direct analog simulation by creating a real three-dimensional potential for atoms \cite{Barredo_2018, 2ym8-vs82} in Rydberg arrays. An experimental study on Rydbergs realized some smaller platonic solids like tetrahedron, cube, and octahedron \cite{PRXQuantum.3.030305}. Although not icosahedral, these results demonstrate the feasibility of such measurements on current technology. Direct analog simulations may also be realized in physically 2D or 1D planar graphs isomorphic to the polyhedron \cite{grunbaum1967convex}, which may be made using platforms that allow for tunable all-to-all connectivity~\cite{Periwal_2021,b9s1-6r44,doi:10.1126/sciadv.aba4935}. Finally, indirect realization by Trotterization or digital simulation, e.g. on superconducting qubit arrays~\cite{Will_2025,andersen2024}
or with mobile neutral atoms~\cite{Evered_2025}, offers a flexible path to exploring icosahedral spectroscopy on diverse graphs.


{\bf Hamiltonians:} We need to realize Hamiltonians for fundamental quantum many-body problems, like the transverse field Ising model, the XY model and other related Hamiltonians. Fortunately, such interactions are very common across all quantum simulation platforms, either natively or through time-averaging techniques~\cite{RevModPhys.76.1037}, and have been studied intensely, and even specifically on Rydberg arrays, trapped ions and superconduting qubits \cite{Bernien_2017, Ebadi_2021, RevModPhys.93.025001, PhysRevX.5.021027}.

{\bf Spectrum \& Quantum Numbers:} Having discussed the feasibility of creating simulators with the necessary  model Hamiltonian with $I_h$ symmetry,  we turn to the measurements necessary to establish the state-operator correspondence. An essential feature which we need is the low lying spectrum of the Hamiltonian, which we refer to as $H$. Luckily methods to extract Hamiltonian spectra from quantum simulations have been developed extensively in general (see e.g.~\cite{Tilly_2022, Baez_2020}), and specifically on real platforms for both Rydberg atoms and trapped ion platforms \cite{Senko_2014, Roushan_2017, martin2026measuringspectralfunctionsdoped}. 

\begin{figure}
    \centering
    \includegraphics[width=1.0\linewidth]{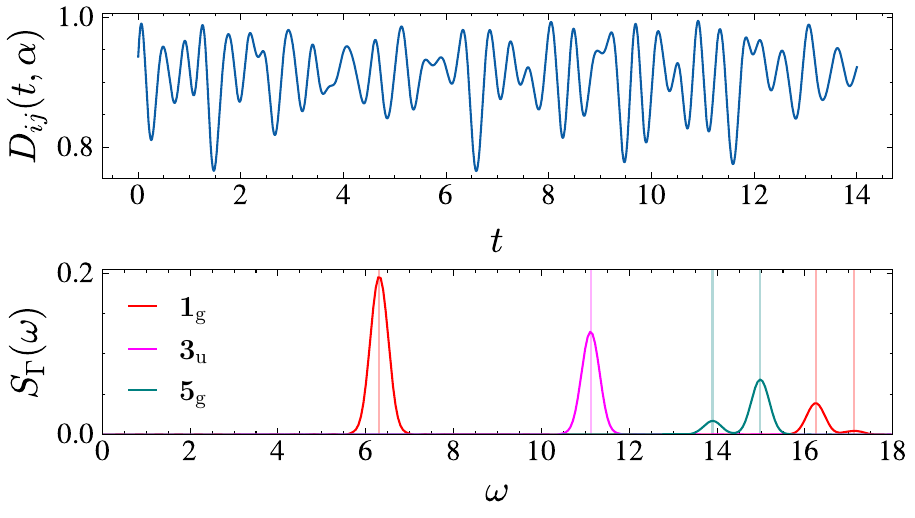}
    \caption{Classical simulation of the quantum simulator protocol we have proposed to extract the spectrum and $I_h$ quantum numbers for icoshedral spectroscopy. (Top) An example of the time series matrix element $D_{ij}(t,\alpha)$ discussed in Eq.~\ref{eq:correlator} for $\mathbb{Z}_{2}$ even operator $\mathcal{O}=X$, $\alpha=\pi / 4$ and separation of graph distance $|i-j|=1$ for qubits  on an icosahedron. We use $(h_\perp,\lambda)=(3.33,0.57)$. (Bottom) $I_{h}$ resolved dynamical structure factor $S_{\Gamma}(\omega)$ computed by processing signal $D_{ij}$ (see Appendix~\ref{app:spectro}). For comparison we have shown as vertical lines the ED energy levels at the same coupling with the same irrep color coding -- they line up well with the peaks in $S_{\Gamma}(\omega)$. Increasing the duration of the quantum simulation sharpens the peaks. To resolve peaks in this case, we need the time series for $T \sim 14$. All energies and times are in units of the Ising exchange constant.}
    \label{fig:time_data}
\end{figure}

In the context of our work, to unambiguously identify the states with operators, we need to  resolve the $I_h$ and \z2sf quantum numbers (like we did in Sec.~\ref{sec:is}). In a real quantum simulation, how this is done will depend in detail on the way the spectrum is extracted. Here we demonstrate one concrete protocol for how the spectrum and quantum numbers can be inferred from an experiment that is natural on quantum simulators. Our idea can be adapted to the platform-specific way in which the spectra is extracted. First, consider the time series matrix elements, 
\begin{equation}
\label{eq:correlator}
    D_{ij}(t,\alpha)=\bra{G}U^{\dagger}_{j,\alpha}e^{+iHt}\mathcal{O}_{i} e^{-iHt} U_{j,\alpha}\ket{G}.
\end{equation}
We argue that this matrix element is naturally accessible in a quantum simulation of Hamiltonian $H$ by interpreting it as the average of the local operator $O_{i}$ in the wavefunction obtained from a quantum simulation protocol: prepare system in the many-body ground state $\ket{G}$ of $H$, act with the single qubit gate $U_{j,\alpha}=e^{-i\alpha \mathcal{O}_j}$, let the state time evolve under $H$ for a time $t$, finally measure the expectation value of the single site obervable $O_i$. We will need the time series function for all inequivalent pairs of $i,j$ and for any one $\alpha$. We show in Appendix~\ref{app:spectro}  that these times series can be post-processed to obtain the spectra and quantum numbers of the excitation. The longer the time series is the more accurate the spectral results are.  To demonstrate its viability, we have carried out a simulation of the entire process for the icosahedral model and verified that we can extract the finite size spectra, Fig~\ref{fig:time_data}.

\section{Discussion \& Conclusions}

In this work, we have proposed how quantum simulators can study the state operator correspondence for $d=3$ CFTs using icosahedral spectroscopy. In Sec.~\ref{sec:is} we studied in detail how this works for the $d=3$ Ising CFT, and demonstrated that by tuning Hamiltonian parameters to satisfy general conformal spectral constraints, many of the light scaling dimensions of the Ising field theory can be extracted with a few percent accuracy from a system of just 20 qubits. We observed that the accuracy of the conformal spectral constraints and of the critical exponents improved with the size of the Hilbert space, going from $2^{12}$ to $2^{20}$, suggesting the accuracy can be further improved by going to a yet larger number of qubits on a quantum computer. 

In Sec.~\ref{sec:qs} we discussed how the proposal's technological requirements are well within the reach of current quantum simulation capabilities. The proposal is clearly amenable to both analog and digital quantum simulation. Of the current physical platforms, Rydberg atoms in optical tweezers seem the most promising analog platform since they already satisfy a majority of the requirements described in Sec.~\ref{sec:qs}. Indeed, an exciting experiment reported very recently has studied the spectrum of the one dimensional Ising chain lattice in Rydberg arrays with up to 35 atoms using modulation spectroscopy and demonstrated the state-operator correspondence to $d=2$ CFTs~\cite{endres2026}. Although the theory behind this experiment is distinct from our work here, with our focus on the challenges specific to $d=3$, the experiment demonstrates the feasibility of a majority of the requirements above in an existing quantum simulation platform.  The promising success of this work makes us optimistic that the study of $d=3$ conformal field theories can be achieved in the near future. The entire process Eq.~\ref{eq:correlator} may also be simulated digitally, on e.g. a trapped ion quantum simulator. 

While we have used the Ising model to illustrate how quantum simulators can extract scaling dimensions, the general method can be extended to a number of different physical problems. For instance, studying truncated $O$($N$) quantum rotors (e.g. spin-1/2 bilayers) on polyhedra gives easy access to the $O$($N$) Wilson-fisher fixed point, including the celebrated XY and Heisenberg universality classes~\cite{sachdev2011}. Strongly interacting fixed points of gauge theories, a topic of great theoretical interest can also be straightforwardly studied using Hamiltonian discretizations such as quantum link models~\cite{Chandrasekharan_1997} in which gauge field qubits are placed on the edges and matter fields are placed on the vertices of polyhedra. Finally, the method outlined above could possibly be extended beyond scaling dimensions to the other important conformal data, the operator product expansion coefficients~\cite{L_uchli_2025}.  All of these ideas are exciting directions that are actively being pursued.

\acknowledgments

We thank B. Gadway, T. Iadecola and G. Murthy for helpful discussions and NSF DMR-2312742 for financial support. This work used  Anvil supercomputer at Purdue University through allocation PHY260072 from the Advanced Cyberinfrastructure Coordination Ecosystem: Services \& Support (ACCESS) program, which is supported by National Science Foundation grants \#2138259, \#2138286, \#2138307, \#2137603, and \#2138296 \cite{Boerner2023ACCESS}.

\appendix

\section{Constraint Optimization}
\label{app:newton}

To find the set of parameters that minimizes a given set of constraints, we implement the Gauss-Newton method as follows. Starting from the current parameter estimate $\theta=(h_\perp,\lambda)$, the constraint vector is linearized as
\begin{equation}
    c(\theta+\delta)\approx c(\theta)+J\delta
\end{equation}
where $J_{ij}=\partial_{j} c_{i}$ is the Jacobian and $\delta$ updates the parameter as $\theta\rightarrow\theta+\delta$. To choose the direction $\delta$, we use the constraint condition $c(\theta+\delta)=0$, which yields the linear system $J\delta = -c(\theta)$. We then solve for $\delta$ via direct inversion if $J$ is square, or more generally, via the pseudoinverse $J^{+}=(J^{T}J)^{-1}J^{T}$, resulting in
\begin{equation}
    \delta = - J^{+}c(x).
\end{equation}
To ensure stability, the update is damped like $\theta\rightarrow \theta+\alpha \delta$ with $\alpha\in(0,1]$, which moves us in a good direction while ensuring the step is not too large. We iterate until $|c|$ is minimized. The Jacobian is computed via the Feynman-Hellman theorem $\partial_{j}E_{n}=\bra{n}\partial_{j}H\ket{n}$, avoiding the need for computing numeric derivatives with finite differences.

\section{Quantum numbers from unequal time two-point correlators}\label{app:spectro}

To use $D_{ij}(t,\alpha)$ (see Eqn.~\ref{eq:correlator}) to resolve the icosahedral quantum numbers, we relate it to the imaginary part of the unequal time two-point correlation function $C_{ij}(t)=\bra{0} \mathcal{O}_{i}(t) \mathcal{O}_{j} \ket{0}$, with $\mathcal{O}\in\{X,Y,Z\}$ a Pauli matrix via
\begin{equation}
\label{eq:D_to_C}
    \mathrm{im}\big(C_{ij}(t)\big)=\frac{D_{ij}(t,
    \alpha)-D_{ij}(t,-\alpha)}{2\sin 2\alpha}.
\end{equation}
The group theoretical projector \cite{tinkham2003group} 
\begin{equation}
\label{eq:proj}
    P_{\Gamma} = \frac{d_{\Gamma}}{|G|}\sum_{g}\chi^{*}_{\Gamma}(g) g
\end{equation}
maps the correlator onto each $I_h$ irrep $\Gamma$ by $C_{\Gamma}(t)=P_{\Gamma}C(t)P_{\Gamma}$. Here $g$ denotes the action on sites as $i\mapsto gi$, and $\chi_{\Gamma}$ denotes the character of the irrep $\Gamma$, which are tabulated. With this, compute the dynamical response $S_{\Gamma}(t)=\mathrm{tr}(\mathrm{im}(C_{\Gamma}(t)))$, and Fourier transform to get the dynamical structure factor $S_{\Gamma}(\omega)$ \cite{Baez_2020, PhysRevB.104.014416}
which is equivalent to 
\begin{equation}
    S_{\Gamma}(\omega)=\pi \sum_{n,i}|\bra{0}\mathcal{O}_{i}\ket{\Gamma,n}|^2\delta(\omega-\omega_{\Gamma,n})
\end{equation}
in the long-time limit, or yields broadened peaks when only a limited amount of time data is available (see Figure~\ref{fig:time_data}).

Using spatial symmetry $C_{ij}=C_{gi,gj}$ one needs to measure only a handful of pair inequivalent correlators and perform the group sum in post-processessing. For example, the icosahedron only has four such pair inequivalent correlators.

\section{Icosidodecahedron}

\begin{figure}[htbp]
\centering
\setlength{\abovecaptionskip}{20pt}
\includegraphics[angle=0,width=0.45\textwidth]{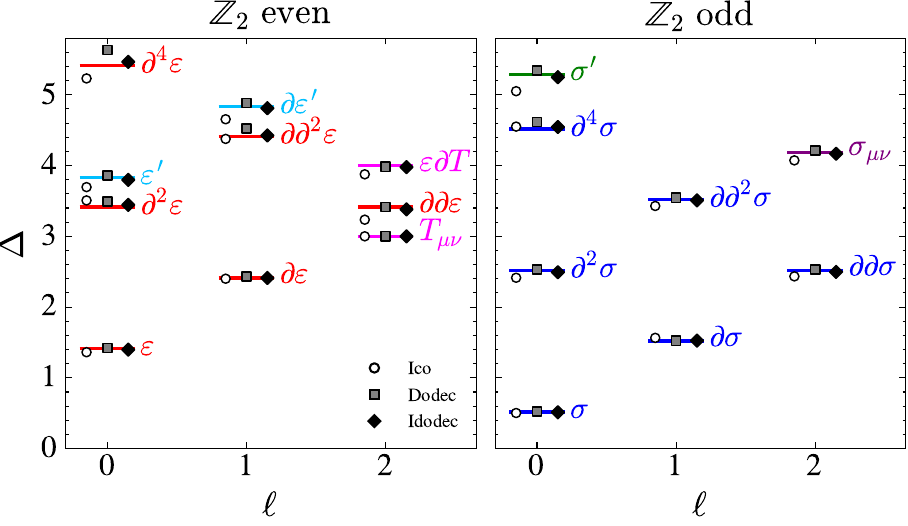}

\caption{Extension of Figure~\ref{spectrum} by the spectrum of the 30-site icosidodecahedron. As with the icosahedron and dodecahedron, the $\sigma$-tower of constraints has been used for the icosidodecahedron. }
\label{fig:spectrum_idodec}
\end{figure}

One additional $I_{h}$-symmetric polyhedron we use to regularize the Ising CFT is the 30-site icosidodecahedron which has site and bond transitivity but lacks face transitivity. The resulting spectrum is shown in Figure~\ref{fig:spectrum_idodec}. We find that the energy levels (scaling dimensions) higher in the energy spectrum including the descendants at higher energies visibly more converged. Unfortuently we cannot provide errors or do scaling analysis. This structure provides more evidence that increasing the number of sites improves the conformal spectrum. Despite this, the scaling dimensions of the lowest primaries themselves do not increase in precision, as reported in Table~\ref{tab:scaling_dimensions_idodec}. 

\label{app:idodec}

\begin{table}[t]
    \centering
    \begin{tabular}{l|ccccc}
    \toprule
    $\mathcal{O}$ & $\sigma$ & $\varepsilon$ & $\varepsilon'$ & $\sigma_{\mu\nu}$ & $\sigma'$ \\
    \midrule
    Bootstrap    & 0.518 & 1.413 & 3.830 & 4.180 & 5.291 \\
    Fuzzy sphere & 0.524 & 1.414 & 3.838 & 4.214 & 5.303 \\
    \lightrule
    Idodec       & 0.513 & 1.398 & 3.797 & 4.167 & 5.248 \\
    Dodec        & 0.526 & 1.420 & 3.860 & 4.210 & 5.342 \\
    Ico          & 0.503 & 1.362 & 3.695 & 4.073 & 5.051 \\
    \bottomrule
    \end{tabular}
        \caption{Extension of Table~\ref{tab:scaling_dimensions} by the conformal dimensions extracted from the icosidodecahedron found by minimizing the $\sigma$-tower of constraints.}
    \label{tab:scaling_dimensions_idodec}
\end{table}

\bibliography{ref}

\end{document}